# Control-theoretic dynamic voltage scaling for embedded controllers

Feng Xia[a,c], Yu-Chu Tian[a], Youxian Sun[b] and Jinxiang Dong[c]

[a]*Faculty of Information Technology, Queensland University of Technology*
*GPO Box 2434, Brisbane QLD 4001, Australia*
[b]*State Key Laboratory of Industrial Control Technology, Zhejiang University*
*Hangzhou 310027, China*
[c]*College of Computer Science and Technology, Zhejiang University*
*Hangzhou 310027, China*
Emails: f.xia@ieee.org; y.tian@qut.edu.au

**Abstract:** For microprocessors used in real-time embedded systems, minimizing power consumption is difficult due to the timing constraints. Dynamic voltage scaling (DVS) has been incorporated into modern microprocessors as a promising technique for exploring the trade-off between energy consumption and system performance. However, it remains a challenge to realize the potential of DVS in unpredictable environments where the system workload cannot be accurately known. Addressing system-level power-aware design for DVS-enabled embedded controllers, this paper establishes an analytical model for the DVS system that encompasses multiple real-time control tasks. From this model, a feedback control based approach to power management is developed to reduce dynamic power consumption while achieving good application performance. With this approach, the unpredictability and variability of task execution times can be attacked. Thanks to the use of feedback control theory, predictable performance of the DVS system is achieved, which is favorable to real-time applications. Extensive simulations are conducted to evaluate the performance of the proposed approach.

## 1 Introduction

With the widespread applications of CMOS integrated circuits, power dissipation has become a critical issue in embedded systems due to the interplay between power consumption, heat dissipation, system reliability and cost [1-3]. In most embedded systems ranging from small handheld devices to large laptop computers, the processor accounts for the major portion of the overall power consumption [4]. Minimizing the power consumption of microprocessors can be performed at different levels of system design, from the circuit and device level (low-level), to the system level (high-level).

Recently, there has been a considerable interest in system-level power-aware design techniques [1]. Among many such techniques, dynamic voltage scaling (DVS), also known as dynamic voltage and frequency scaling, is currently one of the most promising power optimization techniques [4-7]. DVS exploits the convex, normally quadratic, relationship between CPU energy consumption and voltage. By lowering the supply voltage and clock frequency simultaneously, the energy consumption of microprocessors can be reduced quadratically. The majority of existing microprocessors, such as Intel's Xscale and StrongARM, and AMD's K6-2+, support this technique [8].

However, lowering the supply voltage increases the circuit delay. For real-time systems, the supply voltage and clock frequency should be adjusted in a way that all timing constraints are respected [5,6]. In embedded microcontrollers where the performance of control applications is closely related to whether or not the deadlines are met, the system schedulability should be maintained when managing the energy consumption using DVS. Minimising energy consumption and maximising control performance are conflicting, and consequently a fundamental trade-off is required between these two objectives.

Significant effort has been made on DVS mechanisms in many application areas, such as general-purpose computing systems, multimedia, and wireless sensor networks [1,2]. However, limited research has been reported in the literature on feedback control based power management. Varma and colleagues [9] used a PID (Proportional-Integral-Derivative) algorithm to predict the workload for the DVS system. Zhu and Mueller [10] incorporated a PID controller into feedback DVS. Closed-loop DVS algorithms based on the PID con-trol framework have been developed for multimedia systems [11,12]. However, these papers do not exploit control-theoretic design and analysis methodology, i.e., the DVS controllers were not derived from system models.

Soria-Lopezγ et al. [13] presented a proportional control based approach to voltage scaling for soft real-time systems in which a given number of deadline misses are allowed. They have used a simple mathematical model for task scheduling. However, the model was not built for the DVS system. Consequently, the voltage is not determined directly by the feedback controller. Kandasamy et al. [14] have explored a more formal application of control theory in power management. They presented a model predictive control based approach to minimize the energy expenditure of the processor while meeting the quality of service (QoS) requirements of varying workload. The approach was developed for queuing systems. Alimonda et al. [15] developed a control-theoretic approach to feedback DVS for multi-processor system on chip (MPSoC) pipelined architectures. The approach aims to control inter-processor queue occupancy. Wu et al. [16] proposed an analytic approach to DVS in multiple clock domain (MCD) processors. It is based on a dynamic stochastic queuing model and a PI (Proportional-Integral) controller with queue occupancy being the controlled variable.

None of the aforementioned work directly deals with control applications in which the quality of control (QoC) of the target systems is a major concern and depends heavily on real-time execution of control tasks. Simultaneous management of QoC and energy consumption has been studied in [17-21], but no control-theoretic approach for power management has been exploited in these reports.

This paper addresses system-level power management in multitasking microcontrollers that support DVS. The objective is to reduce the CPU energy consumption as much as possible while preserving QoC guarantee. To determine the voltage level of the processor using DVS, the information about task execution times must be gathered. In practice, however, it is hard to obtain (or even estimate) this information, especially when the control algorithm is data dependent or of *anytime* type [22]. This problem is further accentuated in systems where commercial off-the-shelf (COTS) components and non-deterministic operating systems are used. To address the variability and unpredictability of task execution times, a mathematical model will be built in this paper for the DVS system. From this model, a control-theoretic dynamic voltage scaling (ctDVS) scheme will be developed that explores the feedback scheduling methodology [8,23,24]. Thanks to the powerful capacity of feedback control in dealing with uncertainties, the proposed approach can enhance the predictability of the performance of the power manager.

## 2 Problem statement

This section describes the system model and energy consumption model, thus formulating the problem to be addressed in this paper.

### 2.1 System model

Consider an energy-limited variable voltage microprocessor on which $N$ independent control tasks run concurrently. Each control task is responsible for controlling an independent physical process. Assume that the voltage/frequency of the CPU can be adjusted continuously with a scaling factor $\alpha \in [\alpha_{min}, 1]$. Since the clock frequency of CPU is approximately proportional to the supply voltage, $\alpha$ will also be used to denote CPU speed. It is worth mentioning that $\alpha$ is a normalized variable equal to the ratio of actual CPU operating speed to the full speed. For example, for an Intel Xscale processor with a maximum operating voltage of 1.8V, it holds that $\alpha=1.0/1.8=0.556$ when the actual supply voltage is set to 1.0V.

The timing parameters of each control task $i$ are described as follows:
- $h_i$: period, which is equal to the sampling period of the control loop $i$, and is fixed during run time.
- $\hat{c}_{i,nom}$: estimated execution time at full CPU speed. For brevity, it is called *estimated nominal execution time*.
- $\hat{c}_i$: estimated execution time at actual CPU speed associated with $\alpha$, which satisfies $\hat{c}_i = \hat{c}_{i,nom}/\alpha$.
- $c_{i,nom}$: actual execution time at full CPU speed. Assume that $c_{i,nom} = \lambda \hat{c}_{i,nom}$, where $\lambda$ is the *execution time factor*, variable and unpredictable at runtime.

- $c_i$: actual execution time at actual CPU speed, which satisfies $c_i = c_{i,nom}/\alpha = \lambda \hat{c}_i$. It changes with $c_{i,nom}$, and is also unpredictable.

By default, the relative deadline of a control task equals its period under all circumstances. Since practical control applications are usually designed with the capability of tolerating some deadline misses, this paper focuses on soft real-time control tasks. In addition, the following definitions are used:

- CPU *utilization* $U = \sum c_i / h_i$. Accordingly, the estimated CPU utilization $\hat{U} = \sum \hat{c}_i / h_i$.
- CPU *workload* $\omega = U \cdot \alpha = \sum c_{i,nom} / h_i$, and the estimated CPU workload $\hat{\omega} = \hat{U} \cdot \alpha = \sum \hat{c}_{i,nom} / h_i$.

Although different types of real-time task scheduling policies can be employed, we restrict ourselves to illustrate the proposed approach based on the earliest deadline first (EDF) algorithm. According to the well-known schedulable utilization bound for EDF [25], the schedulability condition associated with the processing speed $\alpha$ can be expressed by:

$$\sum_{i=1}^{N} c_i / h_i \leq 1 \Leftrightarrow \omega \leq \alpha \qquad (1)$$

Because $\alpha_{min} \leq \alpha \leq 1$, it is assumed that $\omega = \sum c_{i,nom} / h_i \leq 1$ such that feasible solutions exist under all circumstances. Since the switching time of prevailing processors is always negligibly small in comparison with task periods, the switching overheads including both energy overhead and time overhead between different voltage and frequency levels are neglected.

## 2.2 Energy model

There are basically three components of power dissipation in CMOS circuits: dynamic, static, and short-circuit [1,2]. Among these components, dynamic power contributes to the dominant part of the total power consumption in existing processors. Therefore, this paper targets reducing dynamic power consumption of the CPU. In microcontrollers, the energy expenditure of the processor is sampled at a fixed time interval. As such, the whole CPU energy consumption is related to the energy expenditure per sample as a function of the normalized processing speed $\alpha$. The energy consumption per sample for first-order CMOS delay models is described by [26]:

$$E(\alpha) = CV_0^2 Tf_{max} \alpha \left[ \frac{V_t}{V_0} + \frac{\alpha}{2} + \sqrt{\alpha \frac{V_t}{V_0} + (\frac{\alpha}{2})^2} \right]^2 \qquad (2)$$

where $C$ is the average switched capacitance, $V_t$ is the device threshold voltage, $V_0 = (V_{max} - V_t)^2 / V_{max}$, $T$ is the sampling interval, and $f_{max}$ is the maximum clock speed. For a given DVS system, Sinha and Chandrakasan [27] have shown that (2) can be equivalently approximated by a simple quadratic model:

$$E(\alpha) = \alpha^2 \qquad (3)$$

In this paper, Equation (3) is used to calculate the normalized energy consumption of CPU. It is worth mentioning that our approach will still be applicable if other energy consumption models are used, for example, a more complex model that accounts for both dynamic and static power dissipation. Despite its simplicity, this model has proved illustrative in evaluating the performance of various DVS algorithms [11,12,27]. With this model, it is easy to understand that $\alpha$ should be minimized in order to maximize energy saving. Ideally, the minimum possible CPU speed under task schedulability constraint can be obtained according to (1), which equals $\max\{\omega, \alpha_{min}\}$. In this case, the energy consumption will be minimized if the CPU speed is set to this level. Because $\omega$ is unknown at run-time, however, the exact minimum CPU speed is impossible to deduce as in an ideal case. Therefore, methods are required to handle the uncertainty in task execution times.

## 3 Control-theoretic dynamic voltage scaling

Feedback control theory is one of the most powerful tools for dealing with uncertainties in various engineering systems [22,28,29]. This section aims to develop a control-theoretic dynamic voltage scaling scheme, which we name ctDVS. The DVS system is modelled analytically. A power manager is then designed using feedback control theory. A preliminary stability analysis methodology for the designed DVS system is also given.

*3.1 Basic idea*

Following the idea of feedback scheduling, we propose to treat the DVS system within the microcontroller as a controlled process. The power manager serves as a controller from the viewpoint of control. The choice of some key control-related variables is discussed below.

The *controlled variable* is chosen to be the actual CPU utilization. On one hand, as long as the requested CPU utilization does not exceed the upper bound of the schedulability condition, i.e., 100% for EDF in this paper, all control tasks will be able to complete executions before their deadlines. As a consequence, the QoC will be guaranteed. On the other hand, given that the CPU utilization is controlled at a considerably high level, the idle time of CPU will be reduced, which leads to low energy consumption.

The *manipulated variable* is the CPU speed $\alpha$. This is quite intuitive and is easy to understand, since CPU speed seems to be the only factor that directly determines power dissipation and also effects on control performance. The operating speed of CPU will be adjusted each time the power manager runs, and will remain fixed till the next invocation of the power manager. Similar to general control applications, the purpose of manipulating the CPU speed $\alpha$ is to drive the controlled variable (i.e. CPU utilization) to settle down at a desired level.

The *setpoint* $U_R$ is the desired CPU utilization level. In order for CPU time to be fully utilised and the energy expenditure to be reduced as much as possible, a higher (desired) level of CPU utilization will always be preferable. At the best, the actual CPU utilization will keep exactly at the upper bound of task schedulability condition, i.e. 100%. In real applications, due to the uncertainties in task execution times, the possibility of missing deadlines will increase if $U_R$ approaches 100% too closely. As a result, the control performance may be degraded. On the other hand, if $U_R$ is too low, some resource will be wasted, affecting the effectiveness of energy saving. Therefore, a proper $U_R$ value is often chosen based on knowledge about, e.g., the magnitudes of actual variations of task execution times. In practice, a margin between the setpoint and the schedulable utilization bound will be beneficial to dealing with switching overheads.

Since the power manager is time triggered, with a fixed invocation interval of $T$, the DVS technique employed is naturally interval-based. The system adjusts the operating speed of the processor periodically. During each invocation interval, all tasks run at the same CPU speed. It is worth mentioning that because of the inherent uncertainty and unpredictability of the execution times of tasks, the proposed approach does not provide hard real-time guarantees.

*3.2 Modelling*

As a prerequisite of using feedback control techniques, a mathematical model must be established for the DVS system. For this purpose, examine the following calculation of the CPU utilization in the time interval $[jT, (j+1)T]$:

$$U(j+1) = \sum_{i=1}^{N} \frac{c_i(j)}{h_i} = \lambda(j) \sum_{i=1}^{N} \frac{\hat{c}_i(j)}{h_i} = \frac{\lambda(j)}{\alpha(j)} \sum_{i=1}^{N} \frac{\hat{c}_{i,nom}(j)}{h_i} \qquad (4)$$

where $U(j+1)$ is the output of the DVS system; $\alpha(j)$ is the control input from a control perspective; $\lambda(j)$ is the variable, unknown execution time factor; and the number of control loops $N$ and the estimated nominal execution time $\hat{c}_{i,nom}$ are known yet variable.

Since the estimated execution times of different jobs of a task may be different even in the same DVS invocation interval, the mean of estimated execution times of all jobs associated with each task can be used as $\hat{c}_{i,nom}$ in every interval. For simple description, assume that the estimated execution times of all jobs of a task in the interval $[jT, (j+1)T]$ are equal to $\hat{c}_{i,nom}(j)$.

Because the variability of $\lambda(j)$ could complicate the design of the power manager, a simplification method is used in the modelling. To guarantee stability in all circumstances, the execution time factor $\lambda(j)$ in (4) is replaced by its maximum possible value $K_\lambda = \max\{\lambda(j)\}$. Similar method has been used in modelling CPU task scheduling systems [29]. With $\hat{\omega}(j) = \sum_{i=1}^{N} \frac{\hat{c}_{i,nom}(j)}{h_i}$, Equation (4) can be re-written as:

$$U(j+1) = \frac{K_\lambda \cdot \hat{\omega}(j)}{\alpha(j)} \quad (5)$$

It is seen that the system output $U(j+1)$ has a nonlinear relationship with $\alpha(j)$. To achieve a linear model, let $\beta(j) = 1/\alpha(j)$. Then the following formula is obtained:

$$U(j+1) = K_\lambda \cdot \hat{\omega}(j) \cdot \beta(j) \quad (6)$$

In (6), $\hat{\omega}(j)$ may vary during run time, though it is known. Strictly speaking, this system is a time-variant system. One possible approach to deal with the variability of $\hat{\omega}(j)$ is the same as what we have done with $\lambda(j)$, that is, to use the maximum possible value of $\hat{\omega}(j)$ to replace it. However, unlike $\lambda(j)$ that is unpredictable, $\hat{\omega}(j)$ is known to the system. In this context, an online gain scheduling method [30] is used to compensate for the dynamic variations of $\hat{\omega}(j)$. Accordingly, the term $\hat{\omega}(j)$ is removed from the DVS system model (6). After performing $z$-transform on (6), the following discrete-time model is obtained:

$$G_P(z) = \frac{U(z)}{\Delta\beta(z)} = \frac{K_\lambda}{z-1} \quad (7)$$

where $\Delta\beta(j) = \beta(j) - \beta(j-1)$. Both the CPU utilization $U$ and the variable $\beta$ are subject to saturation, i.e., $0 \leq U \leq 1$, and $1 \leq \beta \leq 1/\alpha_{min}$.

*3.3 Design methodology*

From the viewpoint of feedback control, the model given in (7) is quite simple. In theory, many well-established control techniques can be employed to design the controller, i.e., the power manager. As a simple yet representative illustration, the PI control algorithm is adopted here. The architecture of the DVS loop is shown in Fig. 1. Given below are some reasons why the PI algorithm is adopted:
- The model given in (7) represents a first-order system. It is not hard to design an effective controller for such a system. Therefore, provided that performance requirements are met, the DVS algorithm should be simplified to minimize the runtime overhead.
- PID and variants are the most popular control algorithms in practical control applications. They are well suited for lower-order dynamical systems, and are easy to implement.
- The derivative component of the PID algorithm may amplify the effect of noise, and in consequence is not used. An additional benefit of not using general PID but PI is that this reduces not only the complexity of offline design but also the online computational overhead of the power manager.

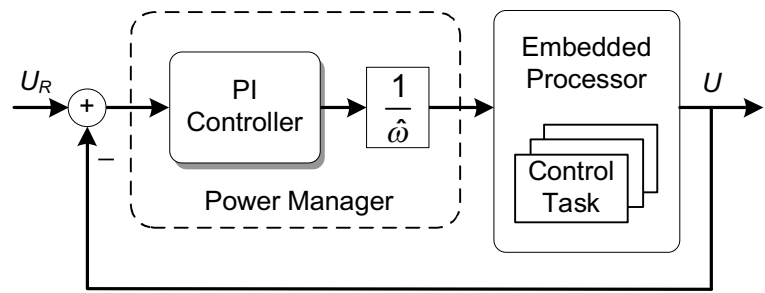

**Fig. 1** *Schematic structure of control-theoretic dynamic voltage scaling*

To determine the coefficients associated with the PI control algorithm, the pole placement method is employed, which is widely used in the control community. In this way, predictable performance of DVS can be achieved explicitly.

The discrete-time transfer function of a PI controller is given by $G_C(z) = \frac{\Delta\beta(z)}{\Delta U(z)} = K_P + K_I \cdot \frac{z}{z-1}$. Combining it with (7) in the framework of Fig. 1 gives the closed-loop transfer function of the DVS loop:

$$G(z) = \frac{G_C(z)G_P(z)}{1+G_C(z)G_P(z)} = \frac{\left(K_P + K_I \cdot \frac{z}{z-1}\right)\frac{K_\lambda}{z-1}}{1+\left(K_P + K_I \cdot \frac{z}{z-1}\right)\frac{K_\lambda}{z-1}} \quad (8)$$

$$= \frac{K_\lambda(K_P + K_I)z - K_\lambda K_P}{z^2 + (K_\lambda K_P + K_\lambda K_I - 2)z + 1 - K_\lambda K_P}$$

Let $a \pm bi$ be the desired closed-loop poles. The corresponding characteristic equation is:

$$(z - a - bi)(z - a + bi) = 0 \quad (9)$$

Rearranging the equation gives:

$$z^2 + 2az + a^2 + b^2 = 0. \quad (10)$$

According to the principle of pole placement, the following equation group is obtained from (8) and (10):

$$\begin{cases} K_\lambda K_P + K_\lambda K_I - 2 = 2a \\ 1 - K_\lambda K_P = a^2 + b^2 \end{cases} \quad (11)$$

Once the desired closed-loop poles are chosen, the control coefficients $K_P$ and $K_I$ can be obtained by simply solving (11). Thus the power manager using the PI algorithm can be designed accordingly.

Now that the power manager is designed using feedback control techniques, control theory can also be employed to analyze the resulting performance, such as stability of the DVS loop. Using established results in the field of discrete-time control [28], it is not difficult to understand the following necessary and sufficient condition for the DVS system stability.

*Theorem 1:* A DVS system designed using the above approach is stable if and only if the closed-loop poles $a \pm bi$ fall inside the unit circle on the $z$ plane, i.e.,

$$a^2 + b^2 < 1 \quad (12)$$

Many equivalent theorems in different forms may be obtained by associating (11) with (12). Using the above design method, not only can the stability but also the transient performance of the DVS system can be determined by the locations of the closed-loop poles on $z$-plane. In other words, different but predictable DVS performance can be achieved through choosing different desired closed-loop poles, i.e., $a \pm bi$.

Given below is a simple example to demonstrate briefly how to calculate the PI coefficients.

*Example 1:* It is known that $K_\lambda = 1.5$ in (7). Desired closed-loop poles are $0.3 \pm 0.1i$. Determine coefficients $K_P$ and $K_I$ of the corresponding PI controller.

*Solution:* Substituting $K_\lambda = 1.5$, $a = 0.3$, and $b = 0.1$ into (11) yields:

$$\begin{cases} 1.5K_P + 1.5K_I - 2 = 0.6 \\ 1 - 1.5K_P = 0.1 \end{cases}$$

Solving the above equation group gives $K_P = 0.6$, and $K_I = 1.13$.

The workflow of the DVS scheme is described as follows. During each invocation interval, the system monitors actual CPU utilization. When the power manager is activated at the $j$-th time instant, it samples current CPU utilization $U(j)$, then compares it with the desired level. Based on the difference, $\Delta\beta(j)$ is calculated using the PI algorithm. Once $\alpha$ is computed from $\alpha(j) = \frac{1}{\beta(j)} = \frac{1}{\beta(j-1) + \Delta\beta(j)}$, it will be multiplied by the gain scheduling component $1/\hat{\omega}(j)$. After that, the processor alters its supply voltage and clock frequency accordingly. The pseudo code of this scheme is given below.

```
//U: Actual CPU utilization
// ŵ: Estimated CPU workload
//α: Normalized CPU speed
Control-Theoretic Dynamic Voltage Scaling {
    Measure U and ŵ;
    //Calculate control input
    Compute ΔU←U_R-U;
    Compute Δβ (w.r.t ΔU) using PI algorithm;
    Compute α based on Δβ (and β);
    Rescale α with 1/ŵ;
    //Reassign CPU speed
    IF α_min≤α≤1
        Assign CPU speed at α;
    ELSEIF α>1
        Assign CPU speed at 1;
    ELSE
        Assign CPU speed at α_min;
    END
}
```

Besides the controller parameters $K_P$ and $K_I$, an important design parameter of ctDVS is the invocation interval $T$ of the algorithm. In this context, the invocation interval determines how often the CPU speed will be changed. Since real processors take time and consume energy to switch between different voltage/frequency levels, a small $T$ may yield considerably large switching overheads due to high frequency of speed change. From this perspective, large invocation intervals are preferable. In addition, to obtain the accurate measurements of CPU utilization (i.e. the feedback information), the interval should not be too small. For example, it must be satisfied that $T \geq \max\{h_i\}$. However, a large $T$ will make the system less sensitive to changes in CPU utilization and/or task execution times, which would in turn degrade the performance of ctDVS. In practice, tradeoffs have to be made between these relevant factors in order to determine an appropriate value of $T$. A possible choice for $T$ is the superperiod of the task set. Another simple way to go is to use a value slightly bigger than $\max\{h_i\}$. With such an invocation interval, the system will be sufficiently sensitive to execution time variations, while incurring only negligible overheads.

## 4 Simulations

In this section, simulation experiments are conducted to evaluate the performance of the proposed approach. Comparison against several representative schemes will also be given.

*4.1 System setup*

Consider an embedded control system composed of three independent control loops. All controlled processes are inverted pendulums with the same linearized model given by:

$$\dot{x} = \begin{bmatrix} 0 & 1 \\ 100 & 0 \end{bmatrix} x + \begin{bmatrix} 0 \\ 100 \end{bmatrix} u + v(t)$$
$$y = \begin{bmatrix} 1 & 0 \end{bmatrix} x + e(t)$$
(13)

where $v$ and $e$ are sequences of white Gaussian noise with zero mean, and their variances are 0.1 and $10^{-4}$, respectively.

The sampling periods of control loops are given by $h_i$ = 20, 25, and 30ms, respectively. All controllers (in the control loops) are well-designed using LQG (Linear-Quadratic-Gaussian) control algorithm, where the optimization objective function is:

$$J = \int_0^\infty (y^2 + 0.01u^2) dt \quad (14)$$

In each run of the simulations, the following accumulative control cost for each control loop is recorded.

$$J_i(t) = \int_0^t (y_i^2(\tau) + 0.01u_i^2(\tau)) d\tau \quad (15)$$

Intuitively, the larger the value of $J$, the worse the QoC.

Four different system design methods are compared, i.e., three most representative traditional methods in addition to the proposed approach.

- DVS-0: The processor always operates at its full speed, i.e., there is no DVS scheme.
- DVS-1: Traditional DVS scheme based on worst-case execution times (WCETs) of tasks. Because hard real-time is guaranteed in this context, the desired CPU utilization level is set to 100% to make full use of the CPU resource. Accordingly, $\alpha = \sum_{i=1}^{3} \frac{WCET_i}{h_i}$.
- DVS-2: Traditional DVS scheme based on the estimated execution times of tasks. It holds that $\alpha = \sum_{i=1}^{3} \frac{\hat{c}_{i,nom}}{h_i}$.
- ctDVS: The approach proposed in this paper. Some related parameters are set as follows: $U_R = 95\%$, $T = 100$ ms, $K_P = 0.6$, and $K_I = 1.13$ (obtained in Example 1).

The minimum allowable scaling factor $\alpha_{min}$ is set to 0.1, quite a small value, such that its effect on DVS is neglected. The examination of this effect is on purpose left for future work. The changing values of execution time factor $\lambda$ are given in Table 1. The estimated nominal execution time of each control task is $\hat{c}_{i,nom} = 4$ ms ($i = 1, 2, 3$). Accordingly $WCET_i = 6$ ms.

**Table 1: Execution time factors used in simulations**

| Time, s | 0-3 | 3-6 | 6-9 | 9-12 |
|---|---|---|---|---|
| $\lambda$ | 0.8 | 1.0 | 0.5 | 1.5 |

## 4.2 Results and analysis

Since the objective of this paper is to save energy while preserving QoC guarantees, it is intuitive that there are basically two aspects of the system performance, i.e., energy consumption and QoC. Consequently, the simulation results are analyzed below from these two perspectives, respectively.

*4.2.1 Energy consumption:* The normalized CPU energy consumption under different schemes is shown in Fig. 2. Since the energy consumption calculated here is a normalized value, it will be given in the form of percentage hereafter.

With the first scheme DVS-0, the processor always operates at the highest possible voltage level, i.e., $\alpha \equiv 1$. Therefore, the corresponding normalized energy consumption $E(\alpha) \equiv 100\%$. It is clear that the energy consumption is the maximum and there is no capability of saving energy in this case.

Under the scheme of DVS-1, $WCET_i$ and sampling periods are fixed and in consequence $\alpha \equiv 0.74$, $E(\alpha) \equiv 54.8\%$. The *normalized energy saving*, which is defined as $100\% - E(\alpha)$, is 45.2%.

Similarly, when the third scheme DVS-2 is employed, $E(\alpha) \equiv 0.49^2 \times 100\% \equiv 24.0\%$ because $\hat{c}_{i,nom}$ remains constant during run time.

In contrast to the above three schemes, ctDVS leads to CPU energy consumption that varies with $\lambda$. The following properties can be observed from Fig. 2:

- When the execution time factor $\lambda$ remains fixed, the energy consumption will settle down to a specific level after a short transient process.
- The resulting energy consumption in steady state changes with different $\lambda$ values.
- ctDVS is capable of dealing with different types of unpredictable workload variations that are characterized by e.g. abrupt increase and decrease of $\lambda$ values.

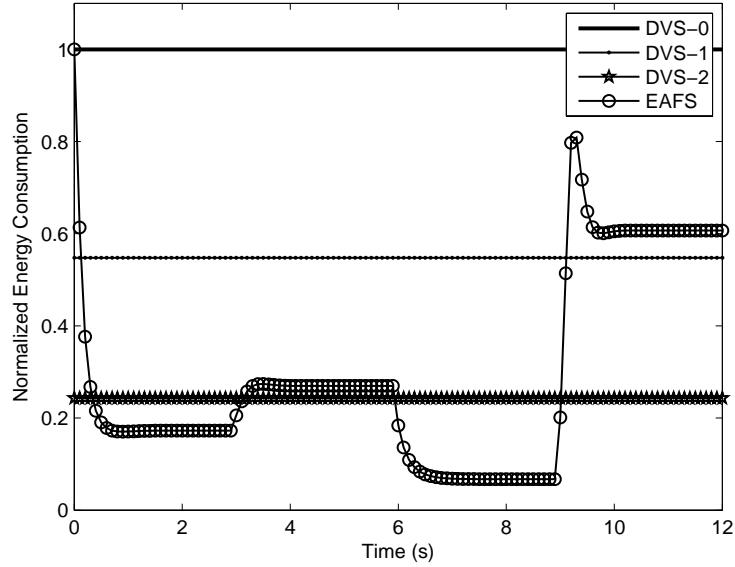

**Fig. 2** *Normalized CPU energy consumption*

Throughout the simulation the average energy consumption in the case of ctDVS comes to 29.7%, which is a little higher than that under DVS-2, but 25.1% lower than that under DVS-1.

From the viewpoint of energy saving, CPU idle time means waste of both computing resource and energy, and hence it should be minimized to maximize the utilization of CPU resources. From this observation, the reasons behind the above results can be explained by the requested CPU utilization of all tasks. Note that requested CPU utilization is not necessarily equal to the actual CPU utilization, because actual CPU utilization is never higher than 100% whereas requested CPU utilization might be.

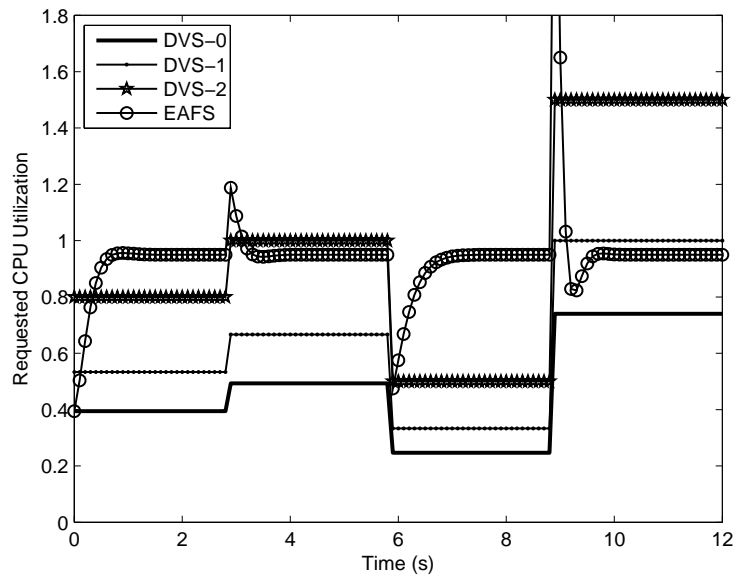

**Fig. 3** *Requested CPU utilization*

As shown in Fig. 3, the (requested) CPU utilization under DVS-0 is always the lowest. For instance, CPU utilization is as low as 25% in the time interval t = 6-8s, which implies severe resource waste. Similarly, the CPU utilization under DVS-1 is also lower than both DVS-2 and ctDVS, and does not exceed the schedulability bound of the EDF algorithm, i.e., 100%. Therefore, the performance of DVS-1 in saving energy is worse than DVS-2 and ctDVS. Under DVS-2, the requested CPU utilization changes with $\lambda$. When $\lambda$ is relatively small, DVS-2 results in significant resource waste. For instance, the CPU utilization under DVS-2 is only 50% when $\lambda = 0.5$. In contrast to dramatic fluctuations of requested CPU utilization under DVS-2, the requested CPU utilization under ctDVS is quite steady. Except for some transient processes, the CPU utilization keeps at the desired high level (i.e. 95%) most of the time. This indicates that CPU time is almost fully used in the case of ctDVS.

It is possible that the system is temporarily overloaded when ctDVS is used. For example, in the short time interval after t = 9s, the abrupt increase of execution time factor from 0.5 to 1.5 causes the requested CPU utilization to be temporarily much higher than the schedulability bound. In this situation some deadlines are missed. When the workload changes frequently, short transient processes are preferable, i.e. the settling time of CPU utilization should be kept sufficiently short so that these changes can be dealt with effectively and in a timely fashion. This can be achieved through well designing the DVS controller. Two possible ways are: 1) to use a short invocation interval $T$; and 2) to tune the controller parameters so that the system can arrive at steady states within a small number of invocation intervals in response to workload variations. Thanks to the use of control theory that leads to predictable performance, the CPU energy consumption and utilization will still act in a similar manner as shown in Figs. 2 and 3, respectively, if the frequency of workload variations increases.

*4.2.2 Quality of control:* Fig. 4 gives the sum of accumulative control costs of three loops, i.e. $\Sigma J_i$. Obviously, all schemes except for DVS-2 achieve satisfactory control performance. The QoC under DVS-0 is the best. Once the DVS technique is introduced under DVS-1, control delays increase slightly, which causes minor degradation in control performance. However, the overall control performance is still comparably good. The performance of ctDVS in terms of QoC is almost identical with that of DVS-1.

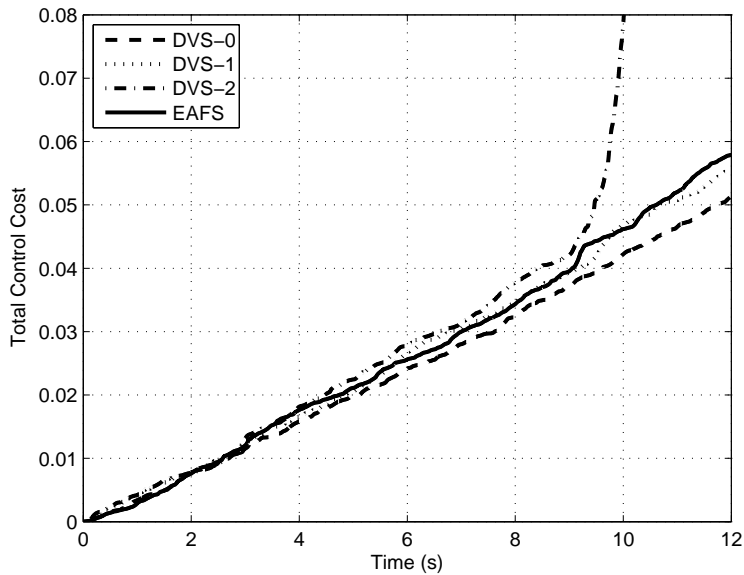

**Fig. 4** *Total control cost of the system*

With DVS-2, the QoC is good until the time instant t = 9s; but the system goes unstable finally. It can be seen from Fig. 3 that the requested CPU utilization increases up to 150% when t > 9s, which is far higher than the schedulability bound of the system. As a consequence, the system is severely overloaded. This is why the system cannot maintain stability.

To summarize, the above simulation results show that:
- In systems where task execution times are unpredictable and time-varying, the proposed ctDVS scheme is capable of not only preserving good control performance, but also reducing remarkably the CPU energy consumption.
- Compared with WCET-based and estimated execution time based traditional DVS schemes, the ctDVS yields much better overall performance.

## 5 Conclusion

This paper deals with power-aware design techniques for embedded microprocessors that run multiple real-time control tasks concurrently. A mathematical model has been deduced for the DVS system. From this model, a control-theoretic design methodology has been developed for DVS-based power managers. The proposed approach is able to tackle the variability and unpredictability of task execution times. Simulation results show that the proposed scheme performs quite well with respect to both energy saving and QoC guarantee in unpredictable environments. In the simulations conducted in this paper, the proposed ctDVS scheme achieves on average 25.1% additional reduction in energy consumption, in comparison with the WCET-based scheme. While compared with the scheme based on estimated execution times, it performs much better in preserving system schedulability, and consequently provides better QoC.

The proposed approach may possibly be extended in several aspects. Firstly, real-life processors generally support only a limited number of voltage/frequency levels. To make ctDVS practically applicable, there is a need for minor extensions, e.g. to bound the obtained scaling factor up to the closest discrete level before voltage adjustment. Secondly, the ctDVS scheme developed in this paper supports only software real-time tasks. An overload handling mechanism may be employed to allow the system to accommodate hard real-time tasks. Thirdly, static power consumption caused by leakage current is expected to increase in the future. In cases where the static power is significant relative to dynamic power, the proposed ctDVS scheme can be combined with a leakage control scheme to reduce both dynamic and static power consumption.

## 6 Acknowledgment

This work is supported in part by Australian Research Council (ARC) under Discovery Projects grant number DP0559111, China Postdoctoral Science Foundation under grant number 20070420232, Australian Government's Department of Education, Science and Training (DEST) under International Science Linkages grant number CH070083, and Natural Science Foundation of China under grant number 60774060.

## 7 References

[1] Unsal, O.S., and Koren, I.: 'System-level power-aware design techniques in real-time systems', *Proceedings of the IEEE*, 2003, **91**, (7), pp.1055-1069
[2] Jha, N.K.: 'Low-power system scheduling, synthesis and displays', *IEE Proc.-Comput. Digit. Tech.*, 2005, **152**, (3), pp.344-352
[3] Andrei, A., Schmitz, M., Eles, P., Peng Z., and Al-Hashimi, B.M.: 'Overhead-conscious voltage selection for dynamic and leakage energy reduction of time-constrained systems', *IEE Proc.-Comput. Digit. Tech.*, 2005, **152**, (1), pp. 28-38
[4] Mao, J.F., Cassandras, C.G., Zhao, Q.C.: 'Optimal Dynamic Voltage Scaling in Energy-Limited Nonpreemptive Systems with Real-Time Constraints', *IEEE Transactions on Mobile Computing*, 2007, **6**, (6), pp. 678-688
[5] Aydin, H., Devadas, V., and Zhu, D.K.: 'System-level Energy Management for Periodic Real-Time Tasks'. Proc. 27th IEEE RTSS, Rio de Janeiro, Brazil, Dec. 2006.
[6] Pillai, P., and Shin, K.G.: 'Real-Time Dynamic Voltage Scaling for Low Power Embedded Operating Systems'. Proc. 18th ACM SOSP, Banff, Alberta, Canada, 2001, pp. 89-102
[7] Choi, J., and Cha, H.: 'Memory-aware dynamic voltage scaling for multimedia applications', *IEE Proc.-Comput. Digit. Tech.*, 2006, **153**, (2), pp.130-136
[8] Xia, F., Dai, X.H., Wang, X.D., and Sun, Y.X.: 'Feedback Scheduling of Real-Time Control Tasks in Power-Aware Embedded Systems'. Proc. 2nd Int. Conf. on Embedded Software and Systems, Xi'an, China, IEEE CS Press, Dec. 2005, pp. 513-518.
[9] Varma, A., Ganesh, B., Sen, M., Choudhury, S. R., Srinivasan, L., Bruce, J.: 'A control-theoretic approach to dynamic voltage scheduling'. Proc. CASES, Georgia, USA, Nov. 2003, pp.255-266
[10] Zhu, Y., Mueller, F.: 'Feedback EDF Scheduling of Real-Time Tasks Exploiting Dynamic Voltage Scaling', *Real-Time Systems*, 2005, **31**, (1-3), pp. 33-63
[11] Lu, Z.J., Hein, J., Humphrey, M., Stan, M., Lach, J., Skadron, K.: 'Control-Theoretic Dynamic Frequency and Voltage Scaling for Multimedia Workloads'. Proc. CASES, 2002, pp. 156-163

[12] Lu, Z.J.,, Lach, J., Stan, M., Skadron, K.: 'Reducing Multimedia Decode Power using Feedback Control'. Proc. 21st Int. Conf. on Computer Design, 2003, pp. 489-496
[13] Soria-Lopezγ, A., Mejia-Alvarez, P., Cornejo, J.: 'Feedback Scheduling of Power-Aware Soft Real-Time Tasks', Proc. 6th Mexican Int. Conf. on Computer Science, Sept. 2005, pp. 266-273
[14] Kandasamy, N., Abdelwahed, S., Sharp, G., Hayes, J.: 'An Online Control Framework for Designing Self-Optimizing Computing Systems: Application to Power Management', Self-Star Properties in Complex Information Systems, O. Babaoglu et al., (Eds.), Lecture Notes in Computer Science, vol. 3460, Springer-Verlag, 2005, pp.174-189
[15] Alimonda, A., Acquaviva, A., Carta, S., and Pisano, A.: 'A Control Theoretic Approach to Run-Time Energy Optimization of Pipelined Processing in MPSoCs'. Proc. DATE, Munich, Germany, 2006, pp. 876-877
[16] Wu, Q., Juang, P., Martonosi, M., and Clark, D.W.: 'Formal Online Methods for Voltage/Frequency Control in Multiple Clock Domain Microprocessors', Proc. ASPLOS-XI, Boston, MA, October 2004, pp. 248-259
[17] Lee, H. S., Kim, B. K.: 'Dynamic Voltage Scaling for Digital Control System Implementation', *Real-Time Systems*, 2005, **29**, pp. 263-280
[18] Zhao, W.H., and Xia, F.: 'Dynamic Voltage Scaling with Asynchronous Period Adjustment for Embedded Controllers', *Dynamics of Continuous, Discrete and Impulsive Systems - Series B*, 2006, **13**, (S1), pp.514-519
[19] Zhao, W.H., and Xia, F.: 'An Efficient Approach to Energy Saving in Microcontrollers', Proc. Asia-Pacific Computer Systems Architecture Conf., Lecture Notes in Computer Science, 2006, vol. 4186, pp. 595-601
[20] Jin, H., Wang, D.L., Wang, H.A., and Wang, H.: 'Feedback fuzzy-DVS scheduling design of control tasks', *Journal of Supercomputing*, 2007, **41**, (2), pp. 147-162
[21] Xia, F., and Sun, Y.X.: 'An Enhanced Dynamic Voltage Scaling Scheme for Energy-Efficient Embedded Real-Time Control Systems', Proc. Int. Conf. on Computational Science and Its Applications, Lecture Notes in Computer Science, 2006, vol. 3983, pp. 539-548
[22] Årzén, K.-E., Robertsson, A., Henriksson, D., Johansson, M., Hjalmarsson, H., Johansson, K.H.: 'Conclusions of the ARTIST2 Roadmap on Control of Computing Systems', *ACM SIGBED Review*, 2006, **3**, (3), pp. 11-20
[23] Xia, F.: 'Feedback Scheduling of Real-Time Control Systems with Resource Constraints', PhD thesis, Zhejiang University, 2006
[24] Xia, F., and Sun, Y.X.: 'Control-Scheduling Codesign: A Perspective on Integrating Control and Computing', *Dynamics of Continuous, Discrete and Impulsive Systems - Series B*, 2006, **13**, (S1), pp. 1352-1358
[25] Liu, C., and Layland, J.: 'Scheduling Algorithms for Multiprogramming in a Hard Real-Time Environment', *J. ACM*, 1973, **20**, pp.46-61
[26] Gutnik, V., and Chandrakasan, A.P.: 'Embedded Power Supply for Low-Power DSP', *IEEE Trans. on VLSI Systems*, 1997, **5**, (4), pp. 425-435
[27] Sinha, A., and Chandrakasan, A. P.: 'Energy efficient real-time scheduling'. Proc. ICCAD, 2001, pp. 458-463
[28] Hellerstein, J.L., Diao, Y.X., Parekh, S., and Tilbury, D.: 'Feedback Control of Computing Systems' (Wiley-Interscience, 2004)
[29] Lu, C., Stankovic, J.A., Tao, G., Son, S.H.: 'Feedback control real-time scheduling: framework, modeling, and algorithms', *Real-time Systems*, 2002, **23**, (1/2), pp. 85-126
[30] Simon, D., Robert, D., Sename, O.: 'Robust Control/Scheduling Co-Design: Application to Robot Control'. Proc. IEEE RTAS, California, USA, Mar. 2005, pp.118-127